\documentclass[12pt,preprint]{aastex}

\usepackage{aas_macros}
\usepackage{natbib}
\bibpunct{(}{)}{;}{a}{}{,} % to follow A&A style

\newcommand {\lsim}{\mbox{$\:\stackrel{<}{_{\sim}}\:$} }

\def\be{\begin{equation}}
\def\ee{\end{equation}}
\def\bea{\begin{eqnarray}}
\def\eea{\end{eqnarray}}

\def\rhor{\rho_{r}}
\def\rhoro{\rho_{r_{o}}}

\def\rhom{\rho_{m}}
\def\rhomo{\rho_{m_{o}}}
\def\rhol{\rho_{\Lambda}}
\def\rhoc{\rho_{crit}}

\def\ol{\Omega_{\Lambda}}
\def\om{\Omega_{m}}

\def\orad{\Omega_{r}}

\def\olo{\Omega_{\Lambda_{o}}}
\def\omo{\Omega_{m_{o}}}
\def\orado{\Omega_{r_{o}}}

\def\tp{t_{Planck}}
\def\ap{a_{Planck}}

\begin{document}
\title{The Cosmic Coincidence as a Temporal Selection Effect Produced by the Age Distribution of 
Terrestrial Planets in the Universe}
\medskip
\author{Charles H. Lineweaver$^{1}$ \& Chas A. Egan$^{1,2}$ \\
\affil{$^{1}$ Planetary Science Institute, Research School of Astronomy and Astrophysics \& \\ 
Research School of Earth Sciences, Australian National University, Canberra, ACT, Australia}
\affil{$^{2}$ Department of Astrophysics, School of Physics, University of New South Wales,\\
Sydney, NSW 2052, Australia}
%Sydney, NSW 2052,Australia \\
%tel: 61-2-9385-5168\\
%fax  61-2-9385-6060\\
charley@mso.anu.edu.au}

%%%%%%%%%%%%%%%%%%%%%%%%%%%%%%%%%%%%%%%%%%%%%%%%%%%%%%%%%%%%%%%%%%%%%%%%%%%%%%
\begin{abstract}
The energy densities of matter and the vacuum are currently
observed to be of the same order of magnitude:
$(\omo \approx 0.3) \sim (\olo \approx 0.7)$.
%$(\omo \appros 0.26) \sim (\olo \approx 0.74)$  t = 13.8.  r = 0.35
The cosmological window of time during which this occurs is relatively narrow. 
Thus, we are presented with the cosmological coincidence problem: 
Why, just now, do these energy densities happen to be of the same order?
Here we show that
this apparent coincidence can be explained 
as a temporal selection effect produced by the age
distribution of terrestrial planets in the Universe.
We find a large ( $\sim 68 \%$) probability that observations made from terrestrial planets
will result in finding $\om$ at least as close to $\ol$ as we observe today.  
Hence, we, and any observers in the Universe who have evolved on terrestrial planets, 
should not be surprised to find $\omo \sim \olo$.
This result is relatively robust if the time it takes an observer to evolve on a terrestrial
planet is less than $\sim 10$ Gyr.
\end{abstract}
%%%%%%%%%%%%%%%%%%%%%%%%%%%%%%%%%%%%%%%%%%%%%%%%%%%%%%%%%%%%%%%%%%%%%%%%%%%%%%

\section{Is the Cosmic Coincidence Remarkable or Insignificant?}
\label{sec:intro}

\subsection{Dicke's argument}

\citet{Dirac1937} pointed out the near equality of several large fundamental dimensionless numbers of the order $10^{40}$. 
One of these large numbers varied with time since it depended on the age of the Universe. 
Thus there was a limited time during which this near equality would hold.  
Under the assumption that observers could exist at any time during the history of the Universe, 
this large number coincidence could not be explained in the standard cosmology.  This problem
motivated \citet{Dirac1938} and \citet{Jordan1955} to construct an ad hoc new cosmology.
Alternatively, \citet{Dicke1961} proposed that our observations of the Universe could only be 
made during a time interval after carbon had been produced in the Universe and before the 
last stars stop shining.  Dicke concluded that this temporal observational selection 
effect -- even one so loosely delimited -- could explain Dirac's large number coincidence 
without invoking a new cosmology.

%%%%%%%%%%%%%%%%%%%%%%%%   Fig 1   Omegas  roller coaster %%%%%%%%%%%%%%%%%%%%%
\begin{figure*}  %[!h]
\epsscale{0.9}
%\centerline{\psfig{figure=rhohistogram.ps,height=11.0cm,width=15.0cm}}
%BoundingBox: 28 70 594 636
%\plotone{omrhos_allnew_130207.ps}
\plotone{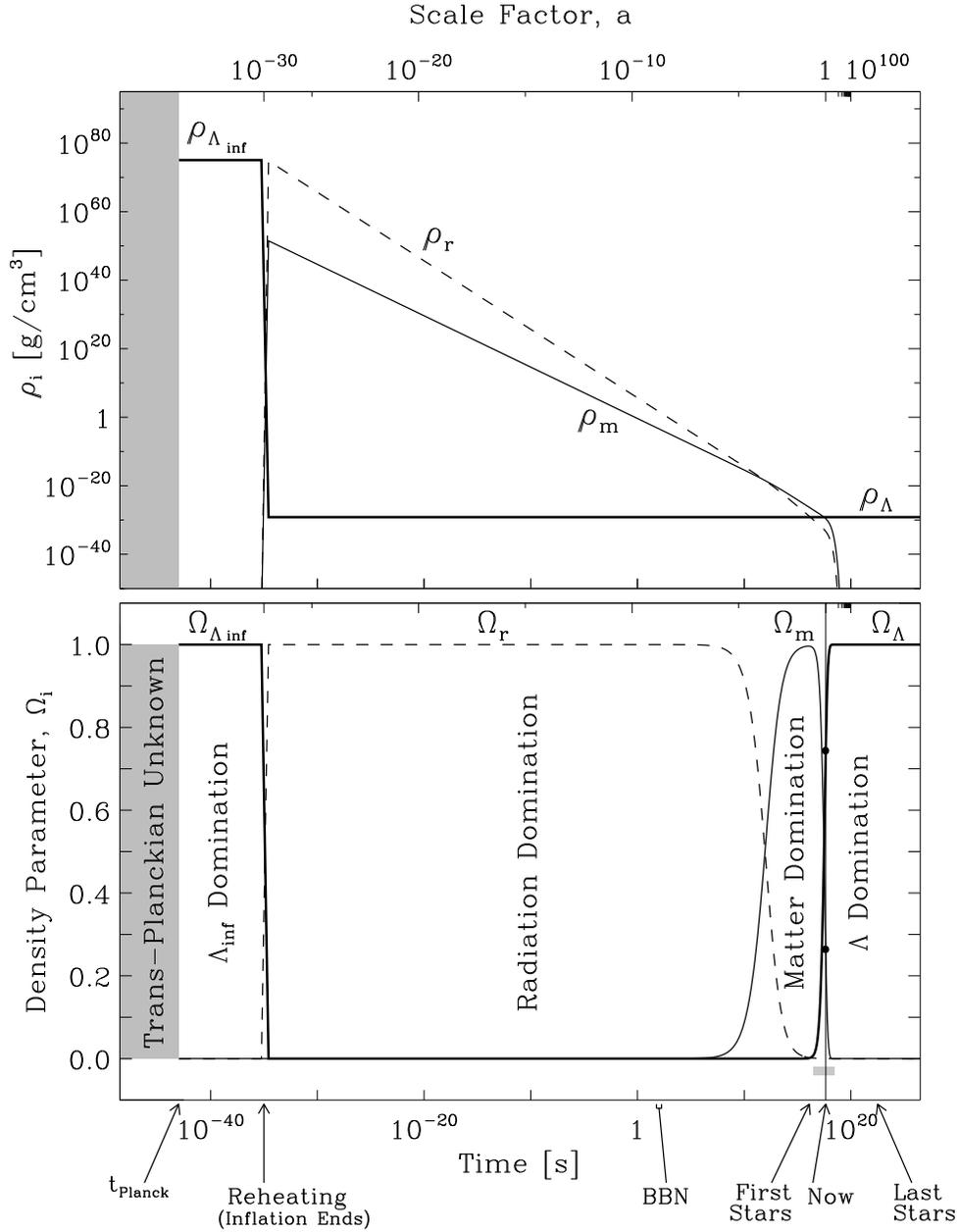}

\caption{The time dependence of the densities of the major components of the Universe.
Given the observed Hubble constant, $H_{o}$ and energy densities in the Universe today,  $\orado$, $\omo$, $\olo$
(radiation, matter and cosmological constant),
we use the Friedmann equation to plot the temporal evolution of the components of the
Universe in g/cm$^{3}$ (top panel), or  normalized to the time-dependent critical 
density $\rhoc = \frac{3 H(t)^{2}}{8\pi G}$ (bottom panel).
We assume an epoch of inflation at $\sim 10^{-35}$ seconds after the big bang and
 a false vacuum energy density $\rho_{\Lambda_{\rm inf}}$ between the Planck 
scale and $t_{GUT}$.
See Table \ref{table:importantimes} and Appendix A for details.
}
\label{fig:omegas}
\end{figure*}
%%%%%%%%%%%%%%%%%%%%%%%%%%%%%%%%%%%%%%%%%%%%%%%%%

Here, we construct a similar argument to address the cosmic coincidence:
Why just now do we find ourselves in the relatively brief interval during which 
$\om \sim \ol$.
The temporal constraints on observers that we present are more empirical and 
specific than those used in Dicke's analysis, but the reasoning is similar.
Our conclusion is also similar:  a temporal observational selection effect can explain 
the apparent cosmic coincidence.  That is,  given the evolution of $\ol$ and $\om$ in our Universe, most observers 
in our Universe who have emerged on terrestrial planets will find $\ol \sim \om$. 
Rather than  being an unusual coincidence, it is what one should expect.

There are two distinct problems associated with the cosmological constant
\citep{Weinberg2000,Garriga2001,Steinhardt2003}. One is the coincidence problem that we address here.
The other is the smallness problem and has to do with the observed
energy density of the vacuum, $\rhol$. Why is $\rhol$ so small compared to the $\sim 10^{120}$ times larger value
predicted by particle physics? Anthropic solutions to this problem invoke a multiverse and
argue that galaxies would not form and there would be no life in a Universe, if $\rhol$ were larger than $\sim 100 $ times its 
observed value \citep{Weinberg1987,Martel1998,Garriga2001,Pogosian2007}. 
Such explanations for the smallness of $\rhol$ do not explain the temporal coincidence 
between the time of our observation and the time of the near-equality of $\om$ and $\ol$.
Here we address this temporal coincidence in our Universe, not the smallness problem in a multiverse.

\subsection{Evolution of the Energy Densities}
\label{sec:subjectivity}

\begin{figure*}[!hpt]
\epsscale{0.8}
%\plotone{r_whynow_logt_loga_shaded_130207.ps}
\plotone{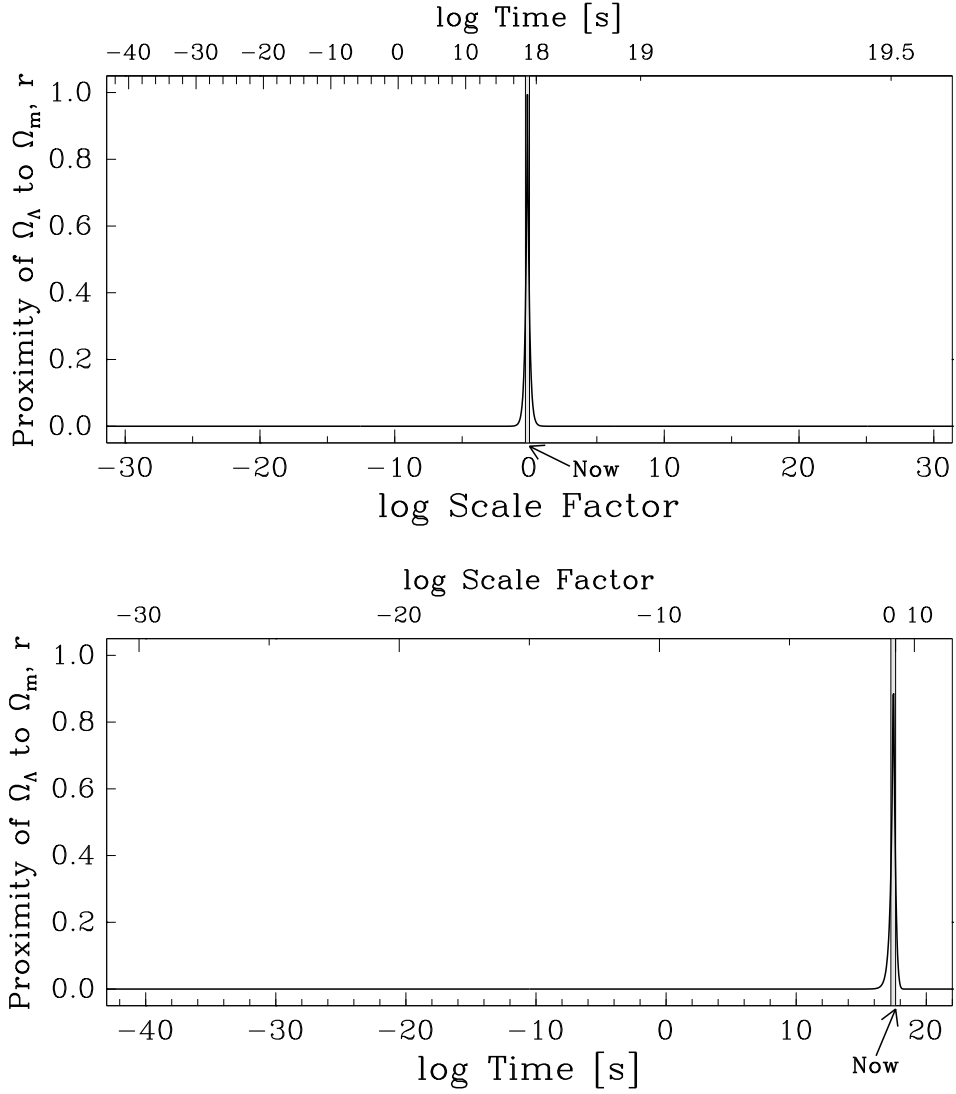}

\caption{
Plot of the  proximity factor $r$ (see Eq. \ref{eq:rdef}).
When the matter and vacuum energy densities of the Universe are the same,  $\om = \ol$, we have $r=1$. 
We currently observe $\omo \sim \olo$ and thus, $r \sim 1$.  
Our existence now when $r \sim 1$ appears to be an unlikely cosmic coincidence 
when the x axis is logarithmic in the 
scale factor (top panel) or logarithmic in time (bottom panel).
In the top panel, following \citet{Carroll2001a}, we have chosen a range of 
scale factors  with ``Now'' midway between the scale factor at the Planck time 
and the scale factor at the inverse Planck time $[\ap < a <  \ap^{-1}]$. 
The brief epoch shown in grey between the thin vertical lines is the epoch during which $r > r_{o}$
(where $r_{o} \approx 0.4$ is the currently observed value).
In the bottom panel the range shown on the x axis is $[\tp < t < 10^{22}]$ seconds.
The Planck time and Planck scale provide reasonably objective lower time limits.
The upper limits are somewhat arbitrary but contribute to the impression that $r \approx 0.4 \sim 1$ 
is an unlikely coincidence.
}
\label{fig:logproblem}
\end{figure*}

Given the currently observed values for $H_{o}$ and the energy densities 
$\orado$, $\omo$ and $\olo$ in the Universe 
\citep{Spergel2006,Seljak2006}, the Friedmann equation tells us the evolution of the scale factor $a$, and the evolution
 of these energy densities.  These are plotted in Fig. \ref{fig:omegas}.
The history of the Universe can be divided chronologically into four distinct periods 
each dominated by a different form of energy: initially the false vacuum energy of inflation dominates, 
then radiation, then matter, and finally vacuum energy. Currently the 
Universe is making the transition from matter domination to vacuum energy domination.
In an expanding Universe, with an initial condition $\om > \ol > 0$, there will be some epoch in which
$\om \sim \ol$, since $\rhom$ is decreasing as $ \propto 1/a^{3}$ while $\rhol$ is a constant (see top panel of
Fig. \ref{fig:omegas} and Appendix A).  
Figure \ref{fig:omegas} also shows that the transition from matter domination 
to vacuum energy domination is occurring now. 
When we view this transition in the context of the time evolution of the Universe (Fig. \ref{fig:logproblem})
we are presented with the cosmic coincidence problem:
Why just now do we find ourselves at the relatively brief interval during which this transition happens?
\citet{Carroll2001a,Carroll2001b} and \citet{Dodelson2000} find this coincidence to be a remarkable result 
that is crucial to understand.
%``It's clearly crucial that we work to better understand this  
%remarkable result.''\citet[p 15, Fig. 3]{Carroll2001a}.
%
The cosmic coincidence problem is often regarded as an important unsolved problem
whose solution may help unravel the nature of dark energy (\citealt{Turner2001,Carroll2001b}).
The coincidence problem is one of the main motivations for the tracker
potentials of quintessence models 
\citep{Caldwell1998,Steinhardt1999,Zlatev1999,Wang2000,Dodelson2000,Armendariz2000,Guo2005}.
In these models the cosmological constant is replaced by a
more generic form of dark energy in which $\om$ and $\ol$ are in
near-equality for extended periods of time.  It is not clear that these models successfully
explain the coincidence without fine-tuning (see \citealt{Weinberg2000,Bludman2004}).

The interpretation of the observation $\omo \sim \olo$ as a remarkable coincidence in need 
of explanation depends on some assumptions that we quantify to determine  
how surprising this apparent coincidence is.
We begin this quantification by introducing a time-dependent proximity parameter, 
\be
r = \min\left[\frac{\ol}{\om},\frac{\om}{\ol}\right]
\label{eq:rdef}
\ee
which is equal to one when $\om = \ol$ and is close to zero when $\om >> \ol$ or $\om << \ol$.
The current value is $r_{o} \approx 0.4$.
In Figure \ref{fig:logproblem} we plot $r$ as a function of log(scale factor) in the upper panel and  
as a function of log(time) in the lower panel.  These logarithmic axes allow a large dynamic range that
makes our existence at a time when $r \sim 1$, appear to be an unlikely coincidence.
This appearance depends on the implicit assumption that we could make cosmological observations 
at any time with equal likelihood. More specifically, the implicit assumption is that 
the {\it a priori} probability distribution $P_{obs}$, of the times we could have made our observations, 
is uniform in log $t$, or log $a$, over the interval shown.  

Our ability to quantify the significance of the coincidence
depends on whether we assume that
$P_{obs}$ is uniform in time, log(time), scale factor or log(scale factor).
That is, our result depends on whether we assume:
$P_{obs}(t) = constant$, $P_{obs}(\log t) = constant$, $P_{obs}(a) = constant$ or 
$P_{obs}(\log a) = constant$. These are the most common possibilities, but there are others.
For a discussion of the relative merits of log and linear time scales and implicit uniform 
priors see Section \ref{sec:measure} and \citet{Jaynes1968}.

\begin{figure*}[!hp]
\epsscale{0.8}
%\plotone{r_whynow_lint_lintshort_shaded_231106.ps}
\plotone{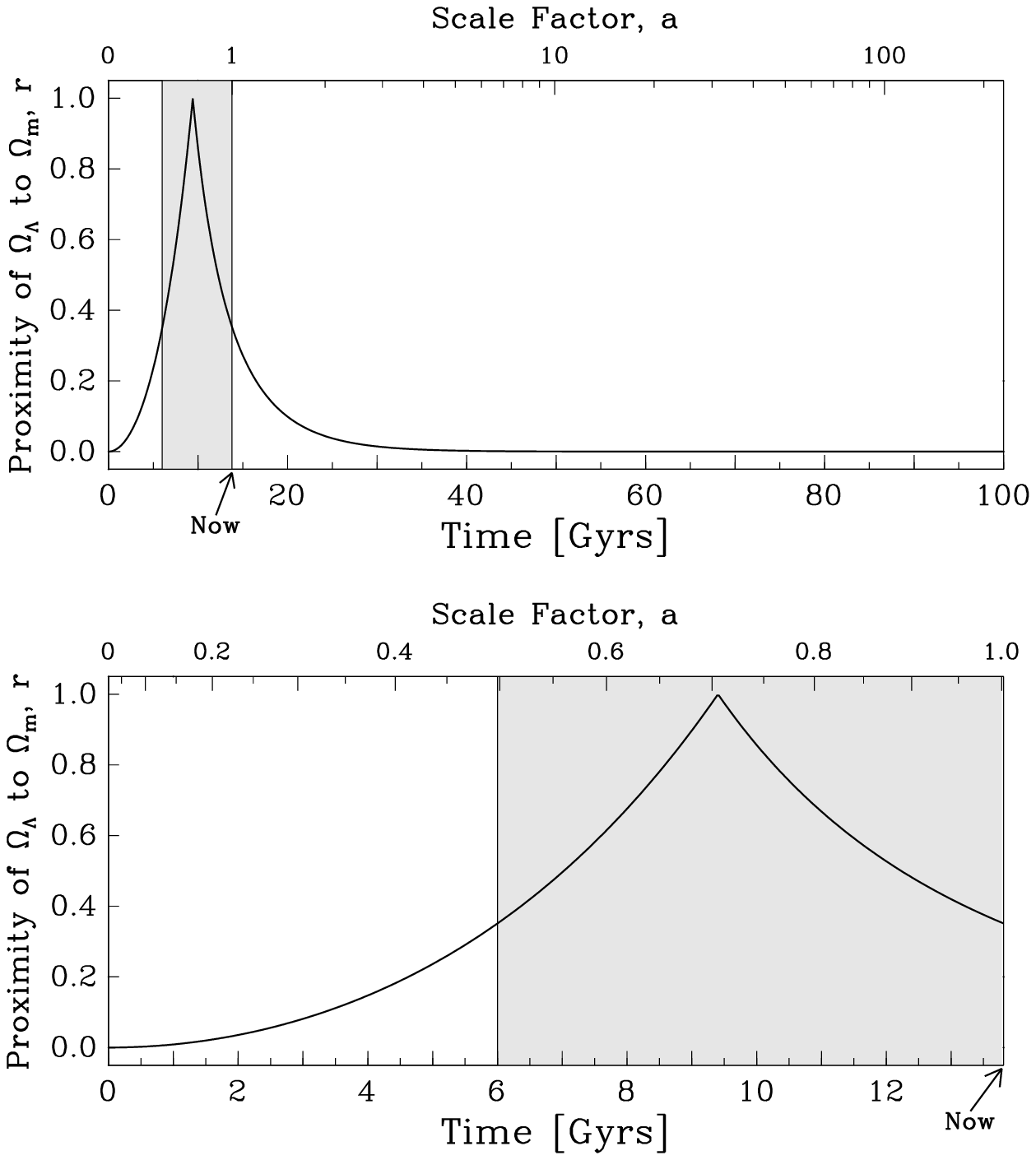}
\caption{
Plot of the  proximity factor $r$, as in the previous figure, but plotted here 
with a linear rather than a logarithmic time axis.  
The condition  $r > r_{o} \approx 0.4$ does not seem as unlikely as in the previous figure.
The range of time plotted also affects this appearance;
with the $[0,100]$ Gyr range of the top panel, the time interval highlighted in grey where $r > r_{o}$,
appears narrow and relatively unlikely.  In contrast, the $[0,13.8]$ Gyr range of the bottom panel
seems to remove the appearance of $r > r_{o}$ being an unlikely coincidence in need of 
explanation;  
for the first $\sim 6$ Gyrs we have
$r < r_{o}$ while in the subsequent $7.8$ Gyr we have $r > r_{o}$. 
How can $r > r_{o}$ be an unlikely coincidence when it has been true for most of the
history of the Universe?
}
\label{fig:linearnoproblem}
\end{figure*}

In Fig. \ref{fig:linearnoproblem} we plot $r(t)$ on an axis linear in time where the 
implicit assumption is that the {\it a priori} probability distribution of our existence is uniform 
in $t$ over the intervals $[0,100]$ Gyr (top panel)
and $[0,13.8]$ Gyr (bottom panel). The bottom panel shows that the observation $r > 0.4$ could have been made 
anytime during the past 7.8 Gyr.  
Thus, our current observation that $r_{o} \approx 0.4$, does not appear to be a remarkable coincidence.
Whether this most recent 7.8 Gyr period is seen 
as ``brief'' (in which case there is an unlikely coincidence in need of explanation) 
or ``long'' (in which case there is no coincidence to explain) depends on whether we view 
the issue in log time (Fig. \ref{fig:logproblem}) or linear time (Fig. \ref{fig:linearnoproblem}).

%\clearpage
%%%%%%%%%%%%%%%%%%%%%%%%  Figure 2  log Coincidence Problem 
%\vspace{15cm}
%%%%%%%%%%%%%%%%%%%%%%%%%%%%%%%%%%%%%%%%%%%%%%%%%
%%%%%%%%%%%%%%%%%%%%%%%%  Figure 3  Linear Coincidence non- Problem %%%%%%%%%%%%%%%%%%%%%%%
%\vspace{15cm}
%%%%%%%%%%%%%%%%%%%%%%%%%%%%%%%%%%%%%%%%%%%%%%%%%%%%%%%%%%%%%%%%%%%%%%%%%%%%%%%%%%%%%%%%%%%%%%%%%

A large dynamic range is necessary to present
the fundamental changes that occurred in the very early Universe, e.g., the transitions at the Planck time, 
inflation, baryogenesis, nucleosynthesis, 
recombination and the formation of the first stars.  Thus a logarithmic time axis is often preferred by 
early Universe cosmologists 
because it seems obvious, from the point of view of fundamental physics, that 
the cosmological clock ticks logarithmically.  
This defensible view and the associated logarithmic axis gives the impression that there is a 
coincidence in need of an explanation.
The linear time axis gives a somewhat different impression.
Evidently, deciding whether a coincidence is of some significance or only an accident is not easy \citep{Peebles1999}.
We conclude that although the importance of the cosmic coincidence problem is subjective, 
it is important enough to merit the analysis we perform here.

The interpretation of the observation $\omo \sim \olo$ as a coincidence in need of explanation 
depends on the {\it a priori}  (not necessarily uniform) probability distribution of our existence. 
That is, it depends on when cosmological observers can exist.
We propose that the cosmic coincidence problem can be more constructively evaluated by replacing these 
uninformed uniform  priors with the more realistic assumption that observers capable of 
measuring cosmological parameters are dependent on the emergence of high density
regions of the Universe called terrestrial planets, which require non-trivial amounts 
of time to form -- and that once these planets are in place, the observers themselves 
require non-trivial amounts of time to evolve.

In this paper we use the age distribution of terrestrial planets estimated by \citet{Lineweaver2001}
to constrain when in the history of the Universe, observers on terrestrial planets 
can exist.
In Section \ref{sec:computeprobability},  %2
we briefly describe this age distribution (Fig. \ref{fig:cosmologists}) and
show how it limits the existence of such observers to an interval in which $\om \sim \ol$ (Fig. \ref{fig:rollercoasterzoom}).
Using this age distribution as a temporal selection function, we compute the probability of an observer on 
a terrestrial planet observing $r \ge r_{o}$ (Fig. \ref{fig:pofr_nonorm}).
In Section \ref{sec:robust} we discuss the robustness of our result and find
(Fig.\ \ref{fig:varyingtev}) that this result is relatively robust
if the time it takes an observer to evolve on a terrestrial planet is less than $\sim 10$ Gyr. 
In Section \ref{sec:discussion} we discuss and summarize our results, and compare it to previous work
to resolve the cosmic coincidence problem \citep{Garriga2000,Bludman2001}.

%%%%%%%%%%%%%%%%%%%%%%%%%%%%%%%%%%%%%%%%%%%%%%%%%%%%%%%%%%%%%%%%
\section{How We Compute the Probability of Observing $\om \sim \ol$}
\label{sec:computeprobability}

\subsection{The Age Distribution of Terrestrial Planets and New Observers}

\begin{figure*}    %[!h!t]
\epsscale{1.0}
%\plotone{pooft231006poft_wmap3_nonorm.ps}
\plotone{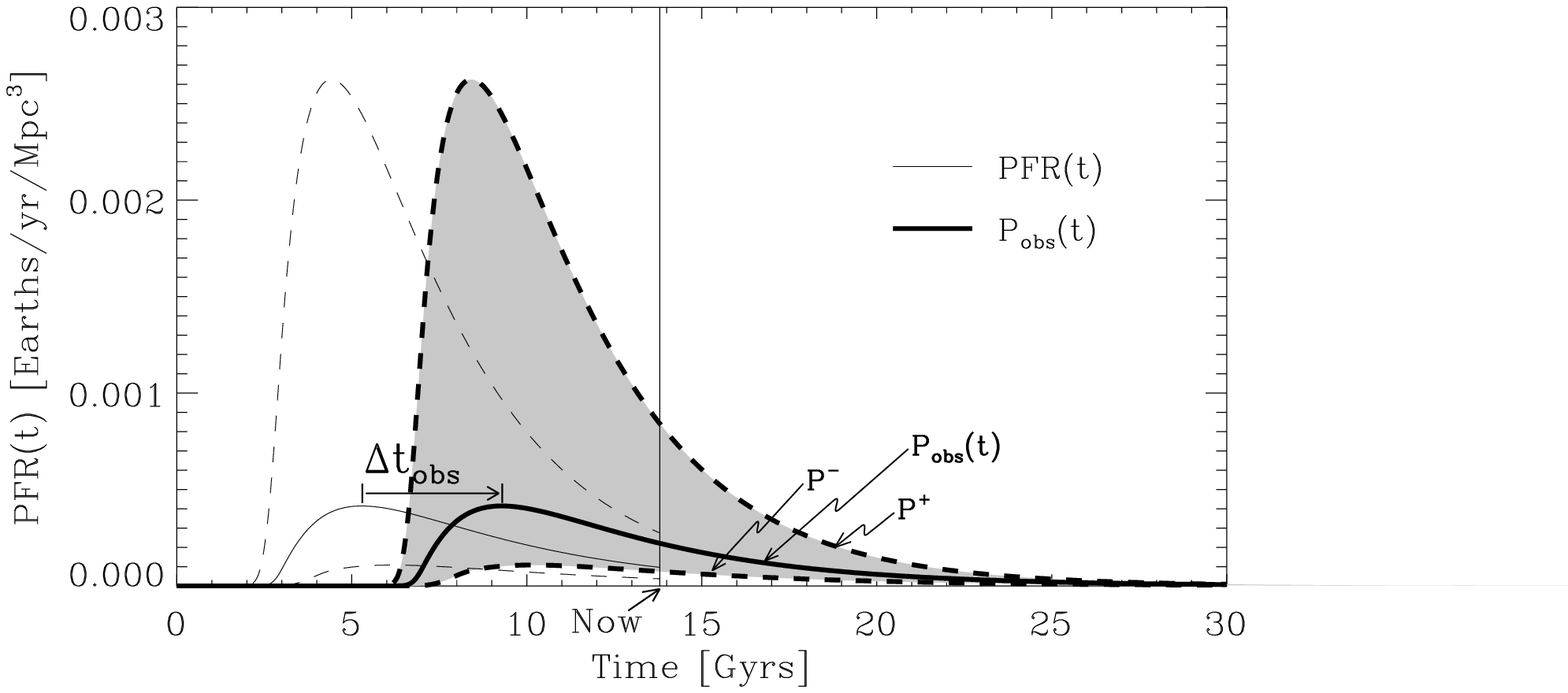}
\caption{The terrestrial planet formation rate $PFR(t)$, derived in \citet{Lineweaver2001} is an 
estimate of the age distribution of terrestrial planets in the Universe and is shown here as a 
thin solid line. Estimated uncertainty is given by the thin dashed lines. %  Note the linear time axis.
To allow time for the evolution of observers on terrestrial planets, we shift this distribution by $\Delta t_{obs}$ 
to obtain an estimate of the age distribution of observers:  
$P_{obs}(t) = PFR(t-\Delta t_{obs})$ (thick solid line). 
The grey band represents the error estimate on $P_{obs}(t)$ which is the shifted error 
estimates on $PFR(t)$.  % from \citet{Lineweaver2001}.
In the case shown here $\Delta t_{obs} = 4$ Gyr, which is how long it took life on Earth to 
emerge, evolve and be able to measure the composition of the Universe.  
To obtain the numerical values on the y axis, we have followed \citet{Lineweaver2001} and assumed 
that one out of one hundred stars is orbited by a terrestrial planet.
We have smoothly extrapolated the $PFR(t)$ of \citet{Lineweaver2001} into the future.
This time dependence and our subsequent analysis does not depend on whether the probability for
terrestrial planets to produce observers is high or low.
}
\label{fig:cosmologists}
\end{figure*}

The mass histogram of detected extrasolar planets
peaks at low masses: $dN/dM \propto M^{-1.7}$, suggesting that low
mass planets are abundant \citep{Lineweaver2003b}.
Terrestrial planet formation may be a common feature of star formation
(\citealt{Wetherill1996a,Chyba1999,Ida2005}).
%Chambers 2006
Whether terrestrial planets are common or rare,  they will have an age distribution proportional to the
star formation rate -- modified by the fact that in the first $\sim 2$ billion years of star formation, 
metallicities are so low that the material for terrestrial 
planet formation will not be readily available.
Using these considerations,  \citet{Lineweaver2001} estimated the age 
distribution of terrestrial planets --  how many Earths are produced by the 
Universe per year, per $Mpc^{3}$ (Figure \ref{fig:cosmologists}). 
%, $PFR(t)$,
If life 
emerges rapidly on terrestrial planets \citep{Lineweaver2002} 
then this age distribution is the age distribution of biogenesis in the Universe. 
However, we are not just interested in any life; we would like to know the distribution in time 
of when independent observers first emerge and are able to measure $\om$ and $\ol$, as we are able to do now.
If life originates and evolves preferentially 
on terrestrial planets, then the \citet{Lineweaver2001} estimate of the age distribution of 
terrestrial planets is an {\it a priori} input which can guide our expectations of when we 
(as members of a hypothetical group of terrestrial-planet-bound observers) 
could have been present in the Universe.
It takes time (if it happens at all) for life to emerge on a new terrestrial planet 
and evolve into cosmologists who can 
observe $\om$ and $\ol$.  
Therefore, to obtain the age distribution of new independent observers
able to measure the composition of the Universe for the first time,
we need to shift the age distribution 
of terrestrial planets by some characteristic time, $\Delta t_{obs}$ required for observers to evolve.
On Earth, it took $\Delta t_{obs} \sim 4 $ Gyr for this to happen.  
Whether this is characteristic of life elsewhere in the Universe is uncertain 
(\citealt{Carter1983,Lineweaver2003a}).
For our initial analysis we use $\Delta t_{obs} = 4$ Gyr as a nominal time to evolve observers.
In Section \ref{sec:deltat} we allow $\Delta t_{obs}$ to vary from 0-12 Gyr to see how sensitive
our result is to these variations.
Fig. \ref{fig:cosmologists} shows the age distribution of terrestrial planet 
formation in the Universe shifted by $\Delta t_{obs} = 4$ Gyr. This curve, labeled ``$P_{obs}$''  
is a crude prior for 
the temporal selection effect of when independent observers can first measure $r$. 
Thus, if the evolution of biological equipment capable of doing cosmology takes about $\Delta t_{obs} \sim 4 $
Gyr, the ``$P_{obs}$'' in Fig. \ref{fig:cosmologists}
shows the age distribution of the first 
cosmologists on terrestrial planets able to look at the Universe and determine the overall 
energy budget, just as we have recently been able to do.

%%%%%%%%%%%%%%%%   Fig 4  shifted age distribution  %%%%%%%%%%%%%%%%%%%%%%
%\vspace{15cm}
%%%%%%%%%%%%%%%%%%%%%%%%%%%%%%%%%%%%%%%%%%%%%%%%%
%\clearpage
%%%%%%%%%%%%%%%%   Figure 5   roller coaster zoom in  %%%%%%%%%%%%%%%%%%%%%%%
%\vspace{15cm}
\begin{figure*}    %[!h!t]
\epsscale{0.9}
%\plotone{zoomomegas241106_wmap3.ps}
\plotone{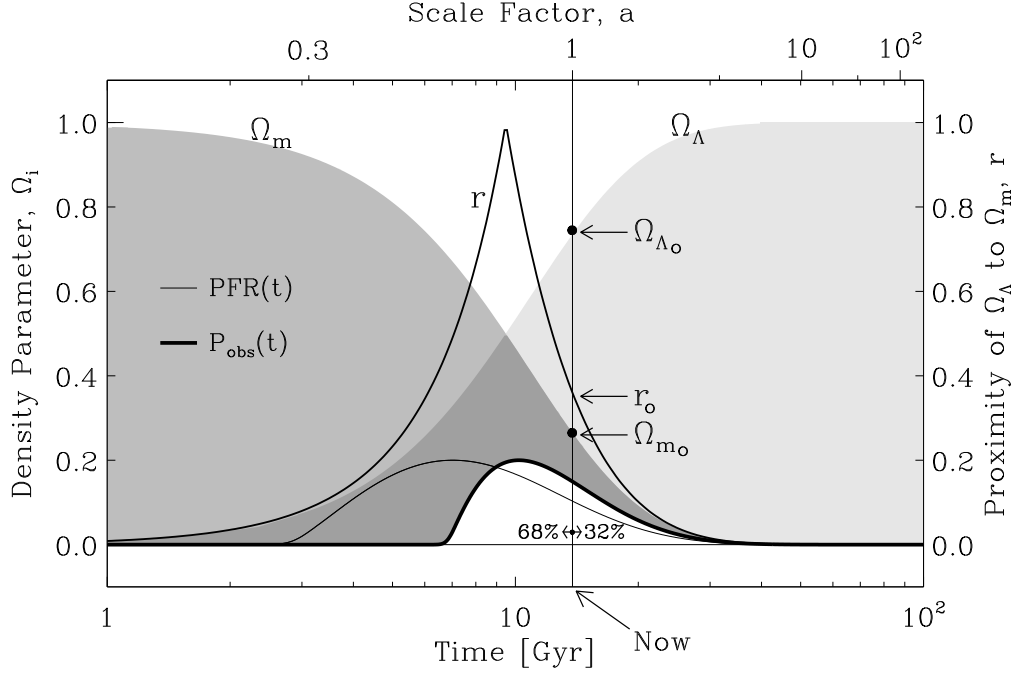}
\caption{
Zoom-in of the portion of Fig. \ref{fig:omegas} between 1 and 100 billion years after
the big bang, containing the relatively narrow window of time in which $\om \sim \ol$. 
The 99 Gyr time interval displayed here is indicated in Fig. \ref{fig:omegas}
by the small grey rectangle above the ``Now'' label.
The proximity parameter $r(t)$ (Eq. \ref{eq:rdef}, Figs. \ref{fig:logproblem} \& \ref{fig:linearnoproblem})
is superimposed.
The thin solid line shows the age distribution of terrestrial planets in the Universe
while the thick solid line is the lateral displacement of this distribution by $\Delta t_{obs} = 4$ Gyr.
These distributions were presented in Fig. \ref{fig:cosmologists}, but here the time axis is logarithmic.
We interpret $P_{obs}$ as the frequency distribution of new observers 
able to measure $\om$ and $\ol$ for the first time.
Since $r(t)$ peaks at about the same time as $P_{obs}(t)$, large values of $r$ 
will be observed more often than small values. 
}
\label{fig:rollercoasterzoom}
\end{figure*}
%%%%%%%%%%%%%%%%%%%%%%%%%%%%%%%%%%%%%%%%%%%%%%%%%

\subsection{The Probability of Observing $\om \sim \ol$.}

In Fig. \ref{fig:rollercoasterzoom} we zoom into the portion of Fig. 
\ref{fig:omegas} containing the relatively narrow window of time in which $\om \sim \ol$.
We plot $r(t)$ to show where $r \sim 1$ and we also plot the age distribution of planets and the 
age distribution of recently emerged cosmologists from Fig. \ref{fig:cosmologists}.
The white area under the thick $P_{obs}(t)$ curve provides an estimate of the time distribution 
of new observers in the Universe.
We interpret $P_{obs}(t)$ as the probability distribution of the times at which new, independent observers 
are able to measure $r$ for the first time. 

\citet{Lineweaver2001} estimated that the Earth is relatively young compared to other terrestrial planets
in the Universe.  It follows under the simple assumptions of our analysis that most 
terrestrial-planet-bound observers will emerge earlier than we have.
We compute the fraction $f$ of observers who have emerged earlier than we have,
\be
f = \frac {\int_{0}^{t_{o}} P_{obs}(t) \; dt}{\int_{0}^{\infty} P_{obs}(t) \; dt} \approx 68\%
\label{eq:older}
\ee
and find that $68\%$ emerge earlier while $32\%$ emerge later.  These numbers are indicated
in Fig. \ref{fig:rollercoasterzoom}.

\subsection{Converting $P_{obs}(t)$  to $P_{obs}(r)$}
\label{sec:convert}
We have an estimate of the distribution in time of observers, $P_{obs}(t)$, and we have
the proximity parameter $r(t)$.  We can then 
convert these to a probability $P_{obs}(r)$, of observed values of $r$. 
That is, we change variables and convert the $t-$dependent
probability to an $r-$dependent probability:  $P_{obs}(t)  \rightarrow P_{obs}(r)$.
We want the probability distribution of the $r$ values first observed by new observers in the Universe. 
Let the probability of observing $r$ in the interval $dr$ be $P_{obs}(r)dr$.  
This is equal to the probability of observing $t$ in the interval $dt$,
which is $P_{obs}(t)dt$ 

\begin{figure*}  [!h!t]
\epsscale{0.7}
%\plotone{pofr_nonorm_wmap3_thatch_130207.ps}
\plotone{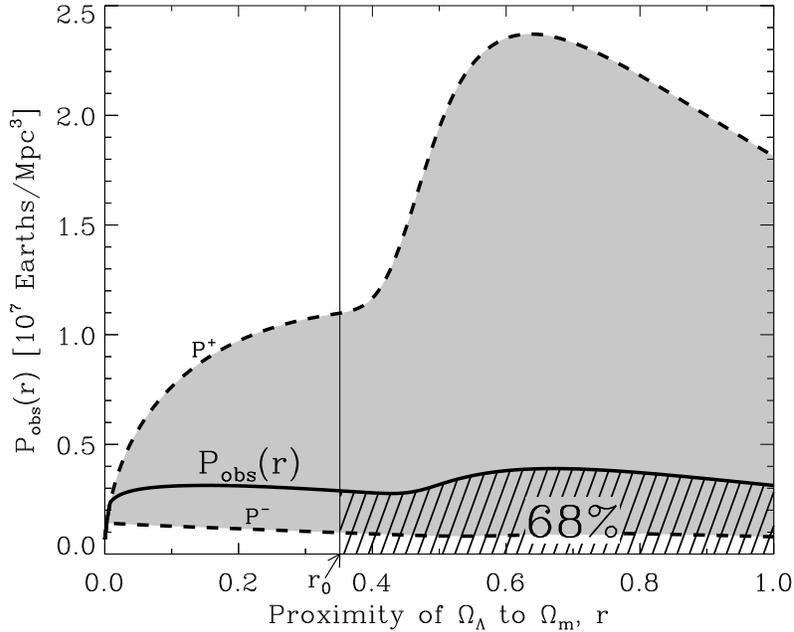}
%%\centerline{\psfig{figure=pofr.ps,height=11.0cm,width=15.0cm}}
%%BoundingBox: 28 70 594 636
\caption{
Probability of new observers on terrestrial planets observing a given 
$r$ (Eq. \ref{eq:prob2}).  % as a function of $r$.
Given our estimate of the age distribution of new cosmologists in the Universe $P_{obs}(t)$,
the probability of observing $\om$ and $\ol$ as close together as
they are, or closer, is the integral given in Eq. (\ref{eq:mainresult}),
shown here as the hashed area labeled $68\%$. 
The dashed lines labeled $P^{+}$ and $P^{-}$ are from replacing $P_{obs}(t)$ 
in Eq. \ref{eq:prob2} with the curves labeled $P^{+}$ and $P^{-}$ 
in Fig. \ref {fig:cosmologists}.
}
\label{fig:pofr_nonorm}
\end{figure*}

Thus,
\be
P_{obs}(r)\;dr = P_{obs}(t)\;dt
\label{eq:prob}
\ee
or equivalently
\be
P_{obs}(r) = \frac{P_{obs}(t)}{dr/dt}
\label{eq:prob2}
\ee
where
$P_{obs}(t) = PFR(t-\Delta t_{obs})$
is the temporally shifted age distribution of terrestrial planets 
and $dr/dt$ is the slope of $r(t)$. Both are shown in Fig. \ref{fig:rollercoasterzoom}.
The distribution $P_{obs}(r)$ is shown in Fig. \ref{fig:pofr_nonorm} along with the upper and lower
confidence limits on $P_{obs}(r)$ obtained by inserting the upper and lower confidence limits
of $P_{obs}(t)$  (denoted ``$P^{+}$'' and ``$P^{-}$'' %thick dashed lines 
in Fig. \ref{fig:cosmologists}), into Eq. \ref{eq:prob2} in
place of $P_{obs}(t)$.

The probability of observing $r > r_{o}$ is,
\be
P(r > r_{o}) = \int_{r_{o}}^{1} P_{obs}(r) \;dr = \int_{t^{\prime}}^{t_{o}} P_{obs}(t) \; dt  \approx 68\%
\label{eq:mainresult}
\ee
where $t^{\prime}$ is the time in the past when $r$ was equal to its present value, i.e., 
$r(t^{\prime}) = r(t_{o}) = r_{o} \approx 0.4$.  We have $t^{\prime} = 6$ Gyr and $t_{o} = 13.8$ Gyr 
(see bottom panel of Fig. \ref{fig:linearnoproblem}).
This integral is shown graphically in Fig. \ref{fig:pofr_nonorm} as the hatched area
underneath the ``$P_{obs}(r)$'' curve, between $r=r_{o}$ and $r = 1$.  
We interpret this as follows: of all observers that have emerged on terrestrial planets, 
68\% will emerge when $r > r_{o}$ and thus will find $r > r_{o}$.
The $68\%$ from Eq. \ref{eq:older} is only the same as the $68\%$ from Eq. \ref{eq:mainresult}
because all observers who emerge earlier than we did, did so more recently than 7.8 billion years
ago and thus, observe $r > r_{o}$  (Fig. \ref{fig:rollercoasterzoom}).

%%\clearpage
%%%%%%%%%%%%%%%%%%%%  Figure 6  probability distributions of r  %%%%%%%%%%%%%%%%%%%%%%%%%%%%%%%
%%\vspace{15cm}
%%%%%%%%%%%%%%%%%%%%%%%%%%%%%%%%%%%%%%%%%%%%%%%%%%

We obtain estimates of the uncertainty on this $68\%$ estimate by computing analogous integrals 
underneath the curves labeled $P^{+}$ and $P^{-}$ in Fig. \ref{fig:pofr_nonorm}.
These yield $82\%$ and $59\%$ respectively.
Thus, under the assumptions made here, $68 ^{+14}_{-10}\%$ of the observers in the Universe will find
$\ol$ and $\om$ even closer to each other than we do.
This suggests that a temporal selection effect due to the constraints on the emergence of
observers on terrestrial planets provides a plausible solution to the cosmic coincidence problem.
If observers in our Universe evolve predominantly on Earth-like planets
(see the ``principle of mediocrity'' in \citet{Vilenkin1995a}), we should not be surprised to find
ourselves on an Earth-like planet and we should not be surprised to find $\olo \sim \omo$.

\section{How Robust is this $68\%$ Result?}
\label{sec:robust}

\subsection{Dependence on the timescale for the evolution of observers}
\label{sec:deltat}

A necessary delay, required for the biological evolution of
observing equipment -- e.g. brains, eyes, telescopes, makes the observation of 
recent biogenesis unobservable 
\citep{Lineweaver2002,Lineweaver2003a}. That is, no observer in the Universe can wake up
to observerhood and find that their planet is only a few hours old.
Thus, the timescale for the evolution of observers, $\Delta t_{obs} > 0$. 

\begin{figure*}    %[!h!t]
\epsscale{0.7}
%\plotone{pofr220806ptev_wmap3_nokey_vals_band.ps}
\plotone{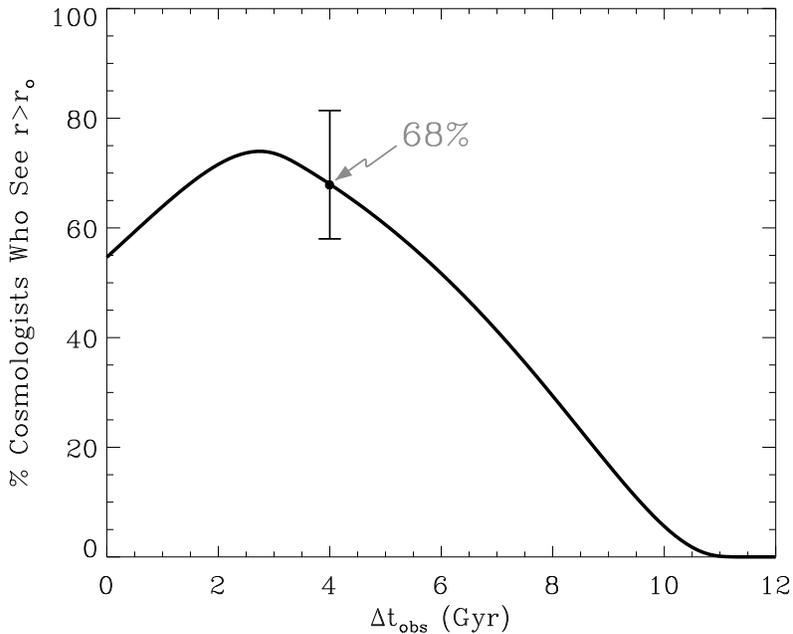}
\caption{Percentage of cosmologists who see $r > r_{o}$ as a function of the time  
$\Delta t_{obs}$, it takes observers
to evolve on a terrestrial planet. Since we have only vague notions about how long
it takes observers to evolve on a planet, we vary $\Delta t_{obs}$ between
0 and 12 billion years and show how the probability  $P(r>r_{o})$ 
 of observing $ r > r_{o}$ (Eq. \ref{eq:mainresult}) varies as a function of $\Delta t_{obs}$.
The $68 ^{+14}_{-10}\%$ point plotted is the result from 
Fig. \ref{fig:pofr_nonorm} where $\Delta t_{obs} = 4$ Gyr.
If $\Delta t_{obs} = 0$,  we use the thin solid line in Fig. \ref{fig:rollercoasterzoom}
as $P_{obs}(t)$ rather than the thick
solid line and we obtain $55\%$. 
}
\label{fig:varyingtev}
\end{figure*}

Our $68 ^{+14}_{-10}\%$ result was calculated under the assumption that 
evolution from a new terrestrial planet to an observer takes $\Delta t_{obs} \sim 4$ Gyr.
To determine how robust our result is to variations in $\Delta t_{obs}$,
we perform the analysis of Sec. \ref{sec:computeprobability}
for  $0 < \Delta t_{obs} < 12$ Gyr. 
The results are shown in Fig. \ref{fig:varyingtev}.  Our $68 ^{+14}_{-10}\%$ result is
the data point plotted at $\Delta t_{obs} = 4$ Gyr.
If life takes $\sim 0$ Gyr to evolve to observerhood, once a terrestrial planet is in place, 
$P_{obs}(t) \approx PFR(t)$
and $55\%$ of new cosmologists would observe an $r$ value larger than the $r_{o} \approx 0.4$
that we actually observe today.
If observers typically take twice as long as we did to evolve ($\Delta t_{obs} \sim 8$ Gyr), there is still a 
large chance ($\sim 30\%$) of observing $r > r_{o}$.  
If $\Delta t_{obs} > 11$ Gyr, $P_{obs}(t)$ in Fig. \ref{fig:rollercoasterzoom}
peaks substantially after $r(t)$ peaks, and the percentage 
of cosmologists who see $r > r_{o}$, is close to zero (Eq. \ref{eq:mainresult}).
Thus, if the characteristic time it takes for life to emerge
and evolve into cosmologists is $\Delta t_{obs} \lsim 10$ Gyr,  our
analysis provides a plausible solution to the cosmic coincidence problem.  

The Sun is more massive than $ 94 \%$ of all stars.  Therefore $94\%$ of stars live longer
than the $t_{\odot}  \approx 10$ Gyr main sequence lifetime of the Sun.  
This is mildly anomalous and it is plausible
that the Sun's mass has been anthropically selected.
For example, perhaps stars as massive as the Sun are needed to provide the UV photons to 
jump start and energize the molecular evolution that leads to life.
If so, then $\sim 10$ Gyr is a rough upper limit to the amount of time a terrestrial planet with simple life has 
to produce observers.  Even if the characteristic time for life to evolve into observers is much longer
than $10 $ Gyr, as concluded by \cite{Carter1983}, this UV requirement that life-hosting stars have 
main sequence lifetimes $\lsim 10$ Gyr would lead to the extinction of most extraterrestrial life
before it can evolve into observers.  This would lead to observers waking to observerhood to find
the age of their planet to be a large fraction of the main sequence lifetime of their star;
the time they took to evolve would satisfy $\Delta t_{obs} \lsim 10 $ Gyr, and they would observe that
 $r \sim 1$ and that other observers are very rare.
Such is our situation.

If we assume that we are typical observers \citep{Vilenkin1995b,Vilenkin1995a,Vilenkin1996a,Vilenkin1996b} and that 
the coincidence problem must be resolved by an observer selection effect \citep{Bostrom2002}, 
then we can conclude that the typical time 
it takes observers to evolve on terrestrial planets is less than $10$ Gyr  ($\Delta t_{obs} < 10$ Gyr).

%\clearpage
%%%%%%%%%%%%%%%%%%%%  Figure 7  varying delta t   probability densities  %%%%%%%%%%%%%%%%%%%%%%%%
%\vspace{15cm}

\subsection{Dependence on the age distribution of terrestrial planets}

The $P_{obs}(t)$ used here (Fig. \ref{fig:rollercoasterzoom})
is based on the star formation rate  (SFR) computed in \citet{Lineweaver2001}.
There is general agreement that the SFR has been declining since redshifts $z \sim 2$.
Current debate centers around whether that decline has only been since $z \sim 2$ or whether the SFR has been declining
from a much higher redshift (\citealt{Lanzetta2002,Hopkins2006,Nagamine2006,Thompson2006}).
Since \citet{Lineweaver2001} assumed a relatively high value for the SFR at redshifts above 2, this
led to a relatively high estimate of the metallicity of the Universe at $z \sim 2$, which corresponds
to a relatively short delay ($\sim 2$ Gyr) between the big bang and the first terrestrial planets.
For the purposes of this analysis, the early-SFR-dependent uncertainty in the $\sim 2 $ Gyr delay is degenerate with, but much
smaller than, the uncertainty of $\Delta t_{obs}$.
Thus the variations of $\Delta t_{obs}$ discussed above subsume the 
SFR-dependent uncertainty in $P_{obs}(t)$.

%%%%%%%%%%%%%%%%%%%%  Figure 8 Naive  Uniform Distributions %%%%%%%%%%%%%%%%%%%%%%%%%%%%%%%%%%%%%%%%%%%%%%
%\vspace{15cm}
\begin{figure*}[!hb]
%\epsscale{0.9} \plotone{multidistrib.ps} 
\epsscale{0.9} 
%\plotone{multidistrib_lite_140207.ps} 
\plotone{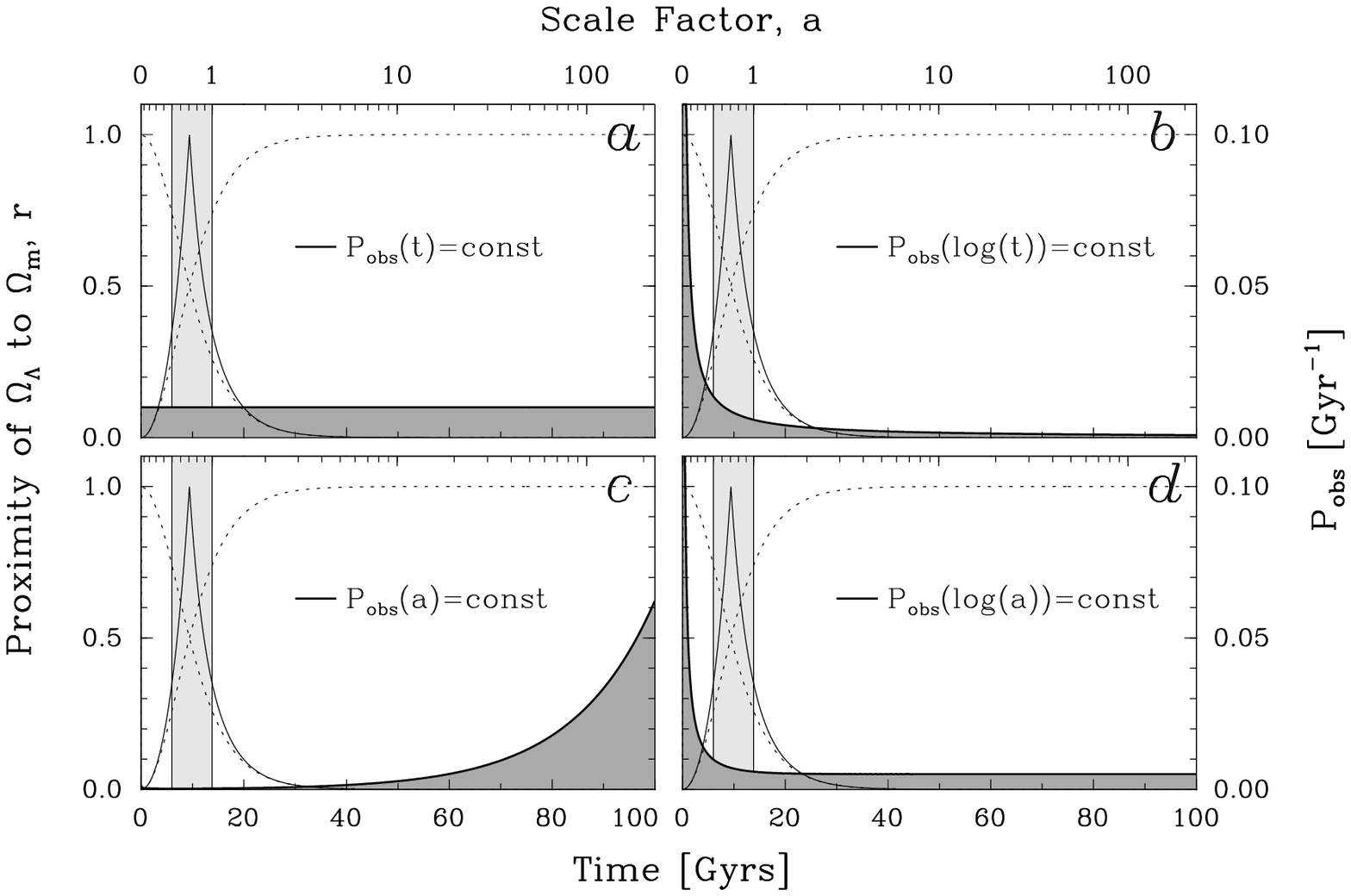} 
\caption{The expected
observed value of $r$ depends strongly on the assumed
distribution of observers over time $t$. This
figure demonstrates a variety of uniform observer distributions $P_{obs}$
which, if used, result in the cosmic coincidence problem that the
observed value of $r$ is unexpectedly high. 
The $P_{obs}$ that are functions
of $log(a)$ or $log(t)$ have been normalized to the interval
$t_{rec}$ to $100$ Gyr.
Panel {\it a}) is the same as the top panel of Fig. \ref{fig:linearnoproblem}.
The probabilities that an observer would fall within the vertical light grey band ($r > r_{o}$)
 in Panels {\it a,b,c} and {\it d} are $8\%,7\%,0.2\%$ and $6\%$
respectively, and are given in the first row of Table \ref{table:minmaxProbs}.
} 
\label{fig:naivedistrib}
\end{figure*}
%%%%%%%%%%%%%%%%%%%%%%%%%%%%%%%%%%%%%%%%%%%%%%%%%

\subsection{Dependence on Measure}
\label{sec:measure}

In Figs. \ref{fig:logproblem}  \& \ref{fig:linearnoproblem}
we illustrated how the importance of the cosmic coincidence depends on the range over 
which one assumes that the observation of $r$ could have occurred.
This involved choosing the range $\Delta x$ shown on the x axis in 
Figs. \ref{fig:logproblem}  \& \ref{fig:linearnoproblem}.
We also showed how the apparent significance of the coincidence depended on how one expressed that range, i.e., 
logarithmic in Fig. \ref{fig:logproblem} and linear in Fig. \ref{fig:linearnoproblem}.  
The coincidence seems most compelling when $\Delta x$ is the largest and
the problem is presented on a logarithmic $x$ axis.  This dependence is  a specific example of a ``measure'' 
problem (\citealt{Aguirre2005,Aguirre2006}).

The measure problem is illustrated in Fig. \ref{fig:naivedistrib},
where we plot four different uniform distributions of observers on a linear time axis.
In Panel {\it a}) $P_{obs}(t)= $ constant. That is, we assume that observers could find themselves
anywhere between $t_{rec} = 380,000$ yr  and $100$ Gyr after the big bang, with uniform probability (dark grey).
In {\it b}), we make the different assumption that observers are distributed uniformly in log(t) over the same
range in time. This means for example that
the probability of finding yourself between 0.1 and 1 Gyr is the same as between 1 and 10 Gyr.
We plot this as a function of linear time and find that the distribution of observers  (dark grey) is highest 
towards earlier times.

To quantify and explore these dependencies further, in Table \ref{table:minmaxProbs}, 
we take the duration when $r > r_{o}$ (call this interval $\Delta x_{r}$)
and divide it by various larger ranges $\Delta x$ (a range of time or scale factor).
Thus, when the probability  $P ( r > r_{o})  = \frac{\Delta x_{r}}{\Delta x}$
is $<< 1$, there is a low probability that one would find oneself in the interval $\Delta x_{r}$ and
the cosmic coincidence is compelling.  However,  when $P( r > r_{o}) \sim 1$ 
the coincidence is not significant.

In the four panels {\it a,b,c} and {\it d} of Fig.  \ref{fig:naivedistrib} the probability of us
observing $r \ge r_{o}$ (finding ourselves in the light grey area) is respectively 
$8\%, 7\%, 0.2\%$ and $6\%$.   These values are given in the first row of  
Table \ref{table:minmaxProbs} along with analogous values when 
11 other ranges for $\Delta x$ are considered.
Probabilities corresponding to the four panels of Figs. \ref{fig:logproblem}  \& \ref{fig:linearnoproblem}
are shown in bold in Table \ref{table:minmaxProbs}.
Our conclusion is that this simple ratio method of measuring the significance of a coincidence yields results 
that can vary by many orders of magnitude depending on the range ($\Delta x$) and 
measure (e.g. linear or logarithmic) chosen.
The use of the non-uniform $P_{obs}(t)$ shown in Fig. \ref{fig:cosmologists}
is not subject to these ambiguities in the choice of range and measure.

\section{Discussion \&  Summary}
\label{sec:discussion}

Anthropic arguments to resolve 
the coincidence problem include \citet{Garriga2000} and \citet{Bludman2001}. 
Both use a semi-analytical formalism (\citealt{Gunn1972,Press1974,Martel1998}) to compute 
the number density of objects that collapse into large galaxies.
This is then used as a measure of the number density of intelligent observers. 
Our work complements these semi-analytic models by using observations of the star formation rate
to constrain the possible times of observation. Our work also extends this previous work by including the 
effect of $\Delta t_{obs}$, the time it takes observers to 
evolve on terrestrial planets. This inclusion puts an important limit on the validity of anthropic solutions
to the coincidence problem. 

\cite{Garriga2000} is probably the work most similar to ours.  They take $\rhol$ 
as a random variable in a multiverse model with a prior probability distribution.  
For a wide range of $\rhol$ (prescribed by a prior based on 
inflation theory) they find approximate equality between the time of galaxy formation $t_{G}$,
the time when $\Lambda$ starts to dominate the energy density of the Universe $t_{\Lambda}$ and now $t_{o}$.
That is, they find that, within one order of magnitude, $t_{G} \sim t_{\Lambda}  \sim t_{o}$. 
Their analysis is more generic but approximate in that it 
addresses the coincidence for a variety of values of $\rhol$  to an order of magnitude precision.
Our analysis is more specific and empirical in that we condition on our Universe and use the \cite{Lineweaver2001}
star-formation-rate-based estimate of the age distribution of terrestrial planets to reach our main result ($68 \%$).

To compare our result to that of \cite{Garriga2000}, we limit their analysis to the $\rhol$ observed in 
our Universe ($\rhol = 6.7 \times 10^{-30} g/cm^{3}$)
and differentiate their cumulative number of galaxies which have assembled up to a given time (their Eq. 9). We
find a broad time-dependent distribution for galaxy formation which is the analog of our more empirical and narrower
(by a factor of 2 or 3) $P_{obs}(t)$.

We have made the most specific anthropic explanation of the cosmic coincidence using the age distribution of
terrestrial planets in our Universe and found this explanation fairly robust to the
largely uncertain time it takes observers to evolve.
Our main result is an understanding of the cosmic coincidence as a
temporal selection effect if observers emerge preferentially
on terrestrial planets in a characteristic time $\Delta t_{obs} < 10 $ Gyr.
Under these plausible conditions,
we, and any observers in the Universe who have evolved on terrestrial planets, 
should not be surprised to find $\omo \sim \olo$.

{\bf Acknowledgements}
We would like to thank Paul Francis and Charles Jenkins for helpful discussions. CE acknowledges a UNSW 
School of Physics post graduate fellowship.\\

%%%%%%%%%%%%%%%%%%%%%%%%%%%%%%%%%%%%%%%%%%%%%%%%%%%%%%%%%%%%%%%%%%%%%%%%%%%%%%%%%%%%%%%%%%%%%%
\newpage
\section*{Appendix A: Evolution of Densities}

Recent cosmological observations have led to the new standard $\Lambda$CDM model in which
the density parameters of radiation, matter and vacuum energy are currently observed to be 
$\orado \approx 4.9 \pm 0.5 \times 10^{-5}$, $\omo \approx 0.26 \pm  0.03$ 
and $\olo \approx 0.74 \pm 0.03$ respectively and
Hubble's constant is $H_{o} = 71 \pm 3 \; km s^{-1} Mpc^{-1}$
\citep{Spergel2006,Seljak2006}.

The energy densities in relativistic particles 
(``radiation'' i.e., photons, neutrinos, hot dark matter), 
non-relativistic particles (``matter'' i.e., baryons,cold dark matter) 
and in vacuum energy scale differently \citep{Peacock1999},

\be
\rho_{i} \propto a^{-3(w_{i}+1)}.
\ee 

Where the different equations of state are, $\rho_{i} = w_{i}\; p$ where $w_{radiation} = 1/3$, $w_{matter} = 0$ 
and $w_{\Lambda} = -1$ \citep{Linder1997}.
That is, as the Universe expands, these different forms of energy density dilute at different rates.

\bea
\rhor & \propto & a^{-4} \\
\rhom & \propto & a^{-3} \\
\rhol & \propto & a^{0} 
\eea

Given the currently observed values for $\orad$, $\om$ and $\ol$, the Friedmann 
equation for a standard flat cosmology tells us the evolution of the scale factor 
of the Universe, and the history of the energy densities:
\bea \label{eq:friedmann}
\left( \frac{\dot{a}}{a} \right)^{2} &=& \frac{8 \pi G}{3} (\rhor + \rhom + \rhol)\\
                                     &=& \frac{8 \pi G}{3}(\rhoro  a^{-4}+ \rhomo a^{-3} + \rhol  a^{0})\\ 
                                     &=& ( \orado a^{-4}+ \omo a^{-3} + \olo  a^{0}) 
\eea
where we have $\rho_{crit} = \frac{3 H(t)^{2}}{8\pi G}$ and $\Omega_{i} = \frac{\rho_{i}}{\rho_{crit}}$.
The upper panel of Fig. \ref{fig:omegas} illustrates these different
dependencies on scale factor and time
in terms of densities while the lower panel shows the corresponding normalized density
parameters. 
A false vacuum energy $\rho_{\Lambda_{inf}}$ is assumed between the Planck scale and the GUT scale. 
In constructing this density plot and setting a value for $\Omega_{\Lambda_{inf}}$ we have used the 
constraint that at the GUT scale, all the energy densities add up to 
$\rho_{\Lambda_{inf}}$ which remains constant at earlier times.

%%%%%%%%%%%%%%%%%%%%%%%%%%%%%%%%%%%%%%%%%%%%%%%%%%%%%%%%%%%%%%%%%%%%%%%%%%%%%%%%%%%%%%%%%%%%%%
%\newpage
\section*{Appendix B: Tables}

%%%%%%%%%%%%%%%%%%%%%%%%%%%%%%%%%%%%%%%%%%%%%%%%%%%%%%%%%%%%%%%%%%%%%%%%%%%%%%%%%%%%%%%%%%%%%%
\begin{table}[!htp]
\begin{center}
\caption{Important Times in the History of the Universe (some used in Table 2)}
\label{table:importantimes}
\scriptsize
\begin{tabular}{|l|c|c|c|}
%\hline
\tableline

Event (ref)                                                   &   Symbol              &      \multicolumn{2}{c}{Time after Big Bang} \vline  \\ 
                                                              &                       &     seconds                 &   Gyr                \\
\tableline
Planck time, beginning of time                                &  $\tp$                &  $5.4 \times 10^{-44}$   &$1.7 \times 10^{-60}$    \\ 
end of inflation, reheating, origin of matter, thermalization &$t_{reheat}$           &  $[10^{-43},10^{-33}]$   &$[10^{-60}, 10^{-50}]$   \\ 
energy scale of Grand Unification Theories (GUT)              &$t_{GUT}$              &  $10^{-33}$              &$10^{-50}$               \\
matter-anti-matter annihilation, baryogenesis                 & $t_{baryogenesis}$    &  $[10^{-33}, 10^{-12}]$  &$[10^{-50},10^{-29}]$    \\    
electromagnetic and weak nuclear forces diverge               & $t_{electroweak}$     &  $10^{-12}$              &$10^{-29}$               \\
light atomic nuclei produced                                  &  $t_{BBN}$            &  $[100, 300]$            &$[3, 9] \times 10^{-15}$ \\
 radiation-matter equality$^1$                                &  $t_{r-m}$            &  $8.9 \times 10^{11}$    &$2.8 \times 10^{-5}$     \\
 recombination$^1$ (first chemistry)                          & $t_{rec}$            &  $1.2 \times 10^{13}$    &$0.38 \times 10^{-3}$    \\
first thermal disequilibrium                                  & $t_{1st therm-dis}$  &  $1.2 \times 10^{13}$    &$0.38 \times 10^{-3}$    \\ 
 first stars, Pop III, reionization$^1$                       & $t_{1st stars}$      &  $1 \times 10^{16}$      &$0.4$                    \\
 first terrestrial planets$^2$                                & $t_{1st Earths}$     &  $8 \times 10^{16}$      &$2.5$                    \\ 
last time $r$ had same value as today                         & $t_{rnow}$           &  $1.9 \times 10^{17}$    &$6.1$                    \\ 
formation of the Sun, Earth$^3$                               & $t_{Sun}$,$t_{Earth}$ &  $2.9 \times 10^{17}$    &$9.1$                    \\
 matter-$\Lambda$ equality$^1$                                & $t_{m-\Lambda}$     &  $3.0 \times 10^{17}$    &$9.4$                    \\
now                                                           & $t_{o}$               &  $4.4 \times 10^{17}$    &$13.8$                   \\
 last stars die$^4$                                           & $t_{last stars}$      &  $10^{22}$               &$10^{6}$                 \\
protons decay$^4$                                             & $t_{proton decay}$    &  $10^{45}$               &$10^{29}$                \\ 
 super massive black holes consume matter$^4$                 & $t_{black holes}$    &  $10^{107}$              &$10^{91}$                \\
 maximum entropy (no gradients to drive life)$^4$             &$t_{heat death}$       &  $10^{207}$              &$10^{191}$               \\
%\hline
\tableline
\end{tabular}
\end{center}
\scriptsize
\noindent
References: \\
(1) \citealt{Spergel2006}, http://map.gsfc.nasa.gov/ \\
(2) \citealt{Lineweaver2001} \\
(3) \citealt{Allegre1995} \\
(4) \citealt{Adams1997} \\
\end{table}
%%%%%%%%%%%%%%%%%%%%%%%%%%%%%%%%%%%%%%%%%%%%%%%%%%%%%%%%%%%%%%%%%%%%%%%%%%%%%%%%%%%%%%

%%%%%%%%%%%%%%%%%%%%%%%%%%%%%%%%%%%%%%%%%%%%%%%%%%%%%%%%%%%%%%%%%%%%%%%%%%%%%%%
\begin{deluxetable}{l l l l l l}
\tablecolumns{6}
\tablewidth{320pt}
\tablecaption{
The probability $P(r > r_{o})$ of observing $r > r_{o}$ assuming a uniform distribution of observers $P_{obs}$
in linear time, log(time), scale factor and log(scale factor) within the range $\Delta x$ listed.
\label{table:minmaxProbs}
}
\tablehead{
\multicolumn{2}{c}{Range  $\Delta x \;^a$}       & \multicolumn{4}{c}{  $P ( r > r_{o})$ [\%]     } \\
\colhead{$x_{min}$} & \colhead{$x_{max}$}    & \colhead{$t$}   & \colhead{$\log(t)$}   & \colhead{$a$} & \colhead{$\log(a)$}
}
\startdata 
%
%$3 \times 10^{15}$& $100$ Gyr      & $0.01\;^b$         & $0.01$                 & $0.10$                & $0.56$ \\ %values from Fig with four assumptions
$t_{rec}$          & $100$ Gyr $^b$   &${\bf 8}\;^c$        & $7$                    & $0.2$                & $6$ \\
$\tp$              & $t_{last stars}$ &$8 \times 10^{-4}$   & ${\bf 0.6}\; ^d$       & $10^{-10^{4}}$        & $10^{-3}$ \\
$\tp$              & $t_{o}$          & ${\bf 60}\;^c$      & $0.6 $                 & $50$                  & $1$ \\ 
$\tp$              & ${\tp^{-1}}$     & $30$                & $0.3$                  & $30$                  & ${\bf 0.5}\;^d$ \\%as per Carroll
$\tp$              & $t_{heat death}$ & $10^{-188}$         & $0.1$                  & $10^{-10^{189}}$      & $10^{-188}$ \\ 
$t_{rec}$          & $t_{protondecay}$& $10^{-26}$         & $1$                     & $10^{-10^{27}}$       & $10^{-26}$ \\ 
$t_{rec}$          &$t_{black holes}$ & $10^{-88}$         & $0.4$                   & $10^{-10^{89}}$       & $10^{-88}$ \\  
$t_{rec}$          & $t_{heat death}$ & $10^{-188}$        &$0.2$                    & $10^{-10^{189}}$      & $10^{-188}$ \\  
$t_{1st stars}$    & $t_{last stars}$ &$8 \times 10^{-4}$  &$6$                      & $10^{-10^{4}}$        & $10^{-3}$ \\
$t_{1st stars}$    & $t_{protondecay}$& $10^{-26}$         & $1$                     & $10^{-10^{27}}$       & $10^{-26}$ \\ 
$t_{1st stars}$    & $t_{black holes}$& $10^{-88}$         & $0.4$                   & $10^{-10^{89}}$       & $10^{-88}$ \\  
$t_{1st stars}$    & $t_{heat death}$ & $10^{-188}$        & $0.2$                   & $10^{-10^{189}}$      & $10^{-188}$ \\ 
\enddata    \\

$^a$ See Table 1 for the times corresponding to columns  1 and 2.\\
$^b$ The four values in the top row correspond to Fig. \ref{fig:naivedistrib}.\\
$^c$ The two values shown in bold in the $t$ column correspond to the two panels of Fig. \ref{fig:linearnoproblem}.\\
$^d$ These values correspond  to the two panels of Fig.\ref{fig:logproblem}.\\ 
\end{deluxetable}

%%%%%%%%%%%%%%%%%%%%%%%%%%%%%%%%%%%%%%%%%%%%%%%%%%%%%%%%%%%%%%%%%%%%%%%%%%%%%%%%%%%%%%%%%%

%\bibliographystyle{agsm}       %{aa}
%\bibliography{anthropic}
\end{document}